\documentstyle[aps,prl,twocolumn,psfig]{revtex}

\begin{document}
\draft                       


\twocolumn[\hsize\textwidth\columnwidth\hsize\csname@twocolumnfalse\endcsname
\title{Numerical investigation of the influence of the history on 
       the local structure of glasses}

\author{Philippe~Jund and R\'emi~Jullien}

\address{Laboratoire des Verres - Universit\'e Montpellier 2\\
         Place E. Bataillon Case 069, 34095 Montpellier France
	}


\maketitle


\begin{abstract}
By means of  molecular dynamics simulations and the Vorono\"\i\
tessellation, we study the influence of the  history on the low 
temperature characteristics of  soft sphere and  silica glasses.
The quench from the liquid is interrupted at an intermediate temperature 
$T_i$ for a given relaxation time, and then the cooling process is 
continued down to 0K.
The local structure at 0K depends on the temperature $T_i$ and the 
effect is larger for $T_i$ close to the glass transition temperature $T_g$. 
This dependence, coherent with recent results, is expected in the 
strong glass former where the characteristics of a particular state 
depend on its history. In the soft-sphere case, because of crystallization 
effects, the dependence of the local structure of the glassy samples on their
history can only be detected in the supercooled liquid region. 
\end{abstract}

\vskip2pc]

\narrowtext

\section*{I. INTRODUCTION }
Experimentally, from the study of slow dynamics in disordered solids,
it is now well established that the time evolution of certain systems
depends on the manner these systems were prepared. This phenomenon called
usually ``aging'' has been observed in structural (polymer) glasses 
\cite{hodge}, spin-glasses \cite{sp} or orientational glasses \cite{levelut}. 
Recently theoretical work concerning aging effects in a structural 
Lennard-Jones glass has been published \cite{kob}. Through classical 
molecular dynamics (MD) simulations the authors show that the dynamical 
behavior of this system depends on the initial temperature before the 
quench and exhibits the same scaling features than the one observed 
in spin-glasses. 
Several theoretical explanations have been proposed to account for
this apparent universal behavior of the out-of-equilibrium dynamics 
in various glasses \cite{bouchaud}.\\
Since it has been shown that aging effects occur in a model glass, 
we want here to address this question from the structural point of view
by using  a combination of classical microcanonical MD calculations 
and the Vorono\"\i\ tessellation. An in-depth description of our Vorono\"\i\ 
tessellation scheme can be found in a previous paper \cite{jundmolsim}.
For the sake of generality, we consider two very different systems, a soft 
sphere glass and a silica glass, representative of amorphous metals and 
vitreous oxides, respectively. 

\section*{II. METHODOLOGY }

The soft sphere system is a very simple 
monoatomic model (without any particular experimental counterpart) described 
by the purely repulsive potential introduced by Laird and Schober 
(LS) \cite{laird}. This potential is basically a modified 
inverse $6^{th}$ power potential similar to those very often used 
in computer simulations of liquids and glasses. Here we use the same 
parameters (arbitrarily equal to 
the Lennard-Jones parameters of Argon) as in previous studies 
\cite{jundmolsim,jundprl} and consequently the glass transition temperature 
$T_g$ and melting temperature $T_m$ are respectively
$T_g \simeq 10.3$K \cite{laird}  and $T_m \simeq 22.7$K \cite{hoover}.
In the present MD calculations we have considered, as before \cite{jundmolsim,jundprl}, 1000 soft spheres in a cubic box of edge length 34.05\AA.
The silica system is described by the more sophisticated potential 
first introduced by van Beest, Kramer and van Santen (BKS) \cite{vanbest} 
and justified by {\it ab initio} calculations.
Though designed originally for the crystalline phases of silica, it has 
been shown to describe quite well the structural \cite{vollmayr}
and vibrational \cite{tarashkin} properties of vitreous silica. In that 
case also we make use of the same parameters as in a previous study \cite{jundphysica} in which we have performed microcanonical MD calculations (treating the Coulomb interactions using the Ewald summation method \cite{ewald}). The
Vorono\"\i\ tessellation procedure has been straightforwardly adapted to this 
two component system and it is worth noting that we did not include any
weighing factor to distinguish silicon from oxygen atoms \cite{jundphysica}. 
The estimated value for the glass transition temperature \cite{vollmayr,tarashkin,jundphysica} is $T_g \simeq 3500$K, subsequently larger than the
experimental one partly because of the very fast cooling rate (2.3 10$^{14}$ K/s) imposed by numerical calculation limitations. But the value of $T_g$ is 
consistent with its dependence with the cooling rate proposed by 
Vollmayr et al. \cite{vollmayr} in which a Vogel-Fulcher dependence of the
relaxation time $\tau^\ast$ of the system on the temperature is assumed. As in \cite{jundphysica}, our silica system 
consists of 216 silicon and 432 oxygen atoms confined
in a cubic box of edge 21.48\AA\ corresponding to a density close to the experimental one. In the following the soft sphere and silica systems
will be called LS and BKS, respectively.

The purpose of the present work is to study the influence of the 
sample history on the low temperature structural properties of our 
two model glasses. To achieve this goal we follow in both cases
a very simple {\em modus operandi}, which is illustrated in Fig. 1 in 
the LS case. We start from well equilibrated liquid samples at a well defined 
temperature $T_\ell$ (point A). We have chosen
$T_\ell = 45$K in the LS case and  7000K in the BKS case.
At this point two paths can be followed : either the liquid is 
quenched to an intermediate temperature $T_i$ (point B) then 
relaxed in a microcanonical ensemble during a time $\tau$ (point C') and 
finally quenched to zero 
temperature (point D) or the liquid is quenched to 0K directly (point C)
relaxed during $\tau$ and finally cooled to 0K (point D). This last 
cooling procedure is necessary because of the slight temperature 
increase following the direct quench to 0K since our 
system evolves at constant energy. From a structural point of view we have checked that this short quench is not really necessary but we have included it anyway in the second path to have a unique final state (point D).

\section*{III. SOFT-SPHERE GLASS}

 In the LS case, the quenching rate $\gamma$ has been chosen equal
to the ``ideal'' quenching rate of $10^{12}$K/s for which the glass 
stability at 0K compared to the crystalline state has been shown to 
be maximum \cite{jundprl} and the local structure at point D has been 
studied using the fraction of pentagonal faces $f_5$
of the Vorono\"\i\ cell attached to each particle. We have previously 
shown that this parameter is a very sensitive tool : a high value of 
$f_5$ (typically 0.45) is a sign of strong icosahedral order, 
characteristic of amorphous metallic glasses, while a small
value of $f_5$ ($<$0.2) indicates the onset of crystal nucleation. 
Indeed, depending on $\tau$, one has to be careful in order to 
distinguish between pure glassy state relaxation and recrystallization effects. Here we have considered two different values of $\tau$ : 100 and 300 ps. 
These values of $\tau$ should be compared to the value of the typical
relaxation time of the system, $\tau^\ast$. This relaxation time
depends on the quenching rate. It is generally admitted that the system falls
out of equilibrium when its relaxation time is on the order of the time scale
of the cooling process i.e. $\tau^\ast \approx T_g/\gamma \approx 
32$ ps where $\gamma$ is the effective cooling rate deduced from Fig.1. The
 values of $\tau$ used in our study are therefore large enough to permit
a complete relaxation of the systems for $T_i \geq T_g$. Of course this is no more the case in the solid phase.\\
Concerning $T_i$, for each sample we saved configurations (positions
and velocities) along the quench at fixed time intervals corresponding 
approximately to a difference of temperature of 2K. The corresponding 
results have then been averaged over ten different samples generated 
from ten independent liquid configurations. It is interesting to note in Fig.1 that the amplitude of the temperature fluctuations during the relaxation
procedure follows the usual $\sqrt T$ behavior and are almost
inexistent at very low temperature. \\
In Fig.2 we show the variation of $f_5$ as a function of $T_i$ for 
$\tau =100$ and $\tau =300$ ps. The error bars show the temperature
dependent dispersion of our statistical ensemble containing ten
samples. 
The first information concerns the high temperature behavior. Fig. 2
points out that if $T_i$ is above a temperature $T_{sup} \approx 19$K 
(lower than $T_m$), the
structure obtained at 0K is the same whichever path the system
followed during the quench. 
This structure is symbolically represented in Fig.2 by the horizontal 
dashed line. Indeed the points for $T_i \geq 19$K lie 
nicely on this line as well as the extrapolated value for $T_i =0$ 
(due to averaging, the value $T_i =0$ can not be reached exactly).
This structure is without any doubt a glassy arrangement and
exhibits a value of $f_5$ close to 0.45 which is a sign of strong
icosahedral local order. This first
information tells us that the low temperature structure following the 
quench will be the same whatever the initial
liquid temperature is, as long as it is above $T_{sup}$. In the picture 
representing the whole system as a point evolving on a complex 
multi-valleyed energy ``surface''\cite{science}, it means that as long 
as the temperature $T_i$ is above $T_{sup}$, the kinetic energy of the system
is greater than all the energy barriers on that surface. Therefore if
$T_i \ge T_{sup}$ the system explores the whole configuration space and the 
low temperature sample does not depend on $T_i$. From a purely practical
point of view it means also that quenches from the liquid state can be started
at temperatures just above the melting temperature without the system
remembering the characteristics of this liquid state.\\
The curves in Fig.2 depart from the horizontal dashed line (which 
represents in a sense the ``reference'' structure) for $T_i$ below
$\approx 19$K. It is 
interesting to note that the two curves
corresponding to the two values of $\tau$ are superimposed 
down to $\approx 15$K. Then these two curves split and exhibit 
a minimum for $T_i \approx 10$K. This minimum is close to 
$0.38$ for $\tau = 100$ps and is much lower for $\tau =300$ps (
$f_5 \approx 0.12$). This is a direct consequence of the relaxation,
the system after $100$ps being on its way to the equilibrium
configuration reached after $300$ps. What is the meaning of this
minimum ? As said earlier a small value of $f_5$ is a sign of
the onset of crystallization. Moreover looking at the results
obtained for the individual samples it appears that some systems
exhibit values of $f_5$ as small as $10^{-2}$ indicating an almost
perfect crystalline character. This can also be verified by direct
visualization of these samples. These results indicate that if
$T_i$ is chosen close to the glass transition temperature $T_g$ 
($T_g = 10.3$K), the propensity to crystallize is maximum. This means that
our model glass exhibits a maximum of instability with respect to
the crystalline state when its temperature is close to $T_g$. Below 
$T_g$, $f_5$ increases again and the low temperature glass stability
is recovered as shown previously for this quenching rate \cite{jundprl}. 
The fact that the tendency to crystallize  increases when
$T_i$ increases up to $T_g$ can be explained
using the same kinds of arguments than the ones developed in \cite{jundprl}. 
When $T_i$ increases, the local density fluctuations increase and 
therefore the probability to find crystal germs in
the samples increases. Of course, as in \cite{jundprl},
 this explanation remains qualitative since it does not take
into account the ability of the existent germs to grow. 
To explain the existence of a minimum around $T_g$ in the $f_5$ curve 
i.e. that the stability increases when $T_i$ increases above $T_g$, 
one can argue that the crystal germ nucleation is now
outweighed by the fact that the energy difference between the 
super-cooled state and the crystal state increases, making the 
crystallization process more difficult to happen. Let us discuss again in terms of relaxation times. When crystal nucleation occurs the nucleation rate has to be taken into account introducing an extra time scale, $\tau_n(T)$. $\tau_n$ 
is a very rapidly varying, non monotonous function of the temperature: it first decreases as T is lowered from the liquid state, due to the decrease in the free energy barrier and then increases again due to the liquid becoming more sluggish. This variation is shown in the so-called ``TTT'' diagrams \cite{zarzy}. The curves in Fig.2 can then be interpreted in the following way: the waiting time
$\tau =100$ps is always smaller than $\tau_n$ while $\tau =300$ps becomes larger than $\tau_n$ between 12.5 and 15K. Hence the samples relaxed at $T_i \leq T_{inf} \approx 13$K exhibit crystal nucleation. 
The maximum crystalline order is reached when
the nucleation rate is maximum i.e. around $T_g$ \cite{zarzy}.

When the curves in Fig. 2 depart from the horizontal dashed line, it
means that the glass ``remembers'' its past history and has 
a different structure depending on the path (ABC'D or ACD) which it
followed during the quench. Nevertheless most of these differences
are due to recrystallization effects which are interesting  but are not 
aging effects (the term aging was in fact introduced to ``distinguish
glassy state relaxation from other time-dependent processes such as
recrystallization''\cite{hodge}). To make this distinction we report
in Fig.3 the variation for $\tau =300$ps of $f_5$ as a function
of $T_i$ (full circles) but also as a function of $T_f$ (open squares). 
$T_f$ is the temperature of the system at point C' in Fig.1. It corresponds 
to the final temperature after the waiting time $\tau$ and can be 
compared directly to the temperature used in the work done on the 
Lennard-Jones glass \cite{kob}. As can be seen in Fig.3, these 
temperatures are almost identical which permits to compare 
the $f_5$s in the system at C' and D. The difference between these 
two states is an additional quench to lower the temperature 
from $T_f$ to 0K. In fact 
the time spent during this quench can be viewed as an extra relaxation
time. Therefore if the system evolves towards a crystalline state
this supplemental relaxation time should lead to a decrease of the
$f_5$s at D compared to the results obtained at C'. In Fig.3 this
is the case for temperatures between 2K and the crossover temperature
$T_{inf}$. This indicates that the departure from the 
reference structure in that temperature range is simply due to 
crystallization effects. In the liquid region for temperatures
above $T_{sup}$ the smaller values of $f_5$ at $T_f$ are a consequence of
thermal effects \cite{euro} and therefore it is expected that once 
the system is cooled down the values of $f_5$ increase towards the values 
obtained in the reference structure. Let us now examine the last
temperature interval which goes from $T_{inf}$ to $T_{sup}$. In
Fig. 3 we see that the additional relaxation time induces an {\em
increase} of the $f_5$s in that temperature region (as an example
at $15$K, $f_5$ is equal to 0.28 at C' while at D $f_5 = 0.42$). 
Therefore no onset of crystallization is present
in the samples. On the contrary the additional quench makes them even 
more amorphous.
Nevertheless in that temperature interval the curve of $f_5$ as a function
of $T_i$ is {\em not} superimposed with the horizontal dashed line. 
Therefore the glassy structures obtained are different from the reference 
structure indicating the effect of the history on the local structure 
of the amorphous phase. The differences are admittedly small but 
they exist and are larger than the error bars in that temperature range. 
This effect exists between $T_{inf}$ and $T_{sup}$ i.e. roughly 
in the supercooled liquid region. It may exist also at lower 
temperatures but to us it  will remain hidden by the recrystallization 
effects. In that temperature region the system does not explore 
the whole configuration space anymore and therefore the glassy 
systems obtained are different from the ones obtained at higher 
temperatures. Locally they appear less disordered 
($f_5$ is smaller than the value obtained for the reference structure)
because the system has not access to the full variety of possible 
configurations but it remains amorphous. In fact we have a picture of 
the configurational space corresponding to $T_i$ in which the 
crystalline ground state is not accessible and which leads to 
glassy systems structurally different from the reference structure. 
This difference is quantified in Fig. 3. 

\section*{IV. SILICA GLASS }

In the BKS case, the quenching rate $\gamma$ has been chosen equal to 
2.3 10$^{14}$K/s and a unique relaxation time $\tau = 42$ps (corresponding 
to 60000 timesteps) has been considered. This choice for $\tau$ results 
from a balance between an upper limit imposed by our computing 
facilities and a lower limit necessary to obtain detectable
effects, even after averaging over ten independent samples, 
as we did here. Also this value of $\tau$ is nearly three times larger than the
typical relaxation time of the system, $\tau^\ast \approx 15$ps, which permits
a complete relaxation of the samples during the waiting time for $T_i \geq T_g$. The difficulty in obtaining these results is rewarded 
by the certitude that in the BKS case no crystallization tendency has 
ever been seen \cite{vollmayr,tarashkin,jundphysica}.
To detect the structural modifications in the case of vitreous silica  
we can no more make use of the $f_5$ parameter since it has a 
significance only for amorphous monoatomic glasses with spherical
symmetry. Here we have chosen to present the standard deviations 
$\sigma_O$ and $\sigma_{Si}$ of the Vorono\"\i\ cell volumes for 
the oxygen and silicon atoms over the whole sample.
In fact we have calculated the dimensionless quantities $\Sigma_O = \sigma_O/<V_O>$ and $\Sigma_{Si} = \sigma_{Si}/<V_{Si}>$ in which the 
standard deviation has been divided by 
the corresponding averaged cell volume. The standard deviation of the cell 
volumes is a direct measure of the local density fluctuations around a 
given atom. As shown in a recent study \cite{jundphysica} this quantity 
decreases when lowering 
the temperature from the liquid state and it is possible to extrapolate it
to zero at 0K as if the system would try to reach a crystalline state.
This decrease slows down  when $T$ becomes lower than the glass 
transition temperature $T_g$, and finally the standard deviations 
saturate to a non zero value at 0K, typical of glassy disorder. 
Therefore the lower the $\sigma$'s are, the closer to an ``ideal'' glassy 
state the considered system is (as long as there is no crystallization).
The variation of $\Sigma$ as a function of $T_i$ is plotted in Fig. 4. 
Note that $\Sigma_{Si}$ is smaller than $\Sigma_O$ which reflects
a larger disorder around the oxygen atoms, in agreement with the fact 
that the tetrahedral arrangement around the silicon atoms is quite well 
preserved, even in the liquid or amorphous states. 
In the case of silica also, the data obtained for a sufficiently large
temperature ($T_i \geq T_{sup} \approx 6000$K) correspond to the  one obtained
for $T_i=0$K (represented by the horizontal dashed lines in the figure) 
within the error bars. The small but significant 
data departure
from the dashed line in the whole range $0<T_i<T_{sup}$ is maximal 
when $T_i$ is 
chosen close to $T_g$, similar to the behavior observed in the soft-sphere
glass. How can this departure be interpreted in the case of silica in which
no crystallization occurs ?
For a strong glass former, at sufficiently low temperature, the characteristics
of a sample depend on the history of the sample (quenching rate, relaxation
time) as shown previously \cite{vollmayr}. Therefore we expect that the local
structure of our silica glass depends on $T_i$ and this is indeed the case
as shown quantitatively in Fig. 4. Nevertheless in that point of view the
shape of the curve needs to be explained: why is there a minimum close to
$T_g$ ? For temperatures below $T_g$, in the solid phase, the structural
relaxation time exceeds by far $40$ps and therefore our waiting time $\tau$ is
not sufficient to permit a complete relaxation of the systems and the situation
gets worse as the temperature decreases. This explains why the curves in Fig.4 converge towards the reference lines at low temperature. At $T_g$, $\tau^\ast 
\approx 15$ps and the waiting time is sufficient to let the systems relax
closer the underlying ideal glassy structure (with a smaller $\Sigma$). In terms of energy landscapes, at $T_g$ the system has enough kinetic energy to overcome the barriers separating the metastable state characteristic of the 
considered quenching rate from the lower lying states, during the waiting
time. At a higher temperature, the system explores more phase space during the time $\tau$ and therefore gets trapped after the quench in a state close
to the metastable state characteristic of the cooling rate. With still
higher temperatures, the system explores the same energy landscape during
the waiting time and the low temperature structure (after the quench) is just
a reflection of the fast cooling rate. 
The minimum in the curves in Fig.4 indicates that 
the structure obtained by stopping
the quench to perform a long relaxation period at a temperature close 
to $T_g$ is closer to the ''ideal'' glassy structure, since it
corresponds to lower values for the standard deviations of the cell 
volumes than in the reference system (quenched at the same speed, but 
without interruption). This is consistent with the fact that, for a 
slower quench, the glass transition temperature would be smaller and so 
would be the standard deviations of the cell volumes 
(which have been shown to decrease with decreasing temperature 
in the liquid phase \cite{jundphysica}). In practice, to obtain the same 
result, it is however more efficient to stop the quench and perform a
relaxation at $T_g$, which is the procedure followed by the authors 
in \cite{tarashkin}, than to perform a slower purely linear quench.

\section*{V. CONCLUSION }

This study combining MD simulations and the Vorono\"\i\ tessellation
in model glasses provides an other point of view in the study 
of the influence of the glass history on its properties. Here we tackle
this question from the structural point of view.
First of all the study as a function of temperature of the local
structure in soft-sphere samples has shown that this  model glass exhibits
a maximum instability with respect to the crystalline state at
$T_g$.\\
We have then shown that the structure of this model glass
depends on its history. Indeed when a system is cooled
to an intermediate temperature, relaxed and then quenched to 
zero temperature (and if it stays amorphous) its structure is 
different from the structure obtained by a direct quench to 0K 
followed by the same relaxation period {\em if} the intermediate 
 temperature is chosen roughly in the supercooled liquid region. 
Secondly we did the same kind of study in a model silica glass where
the crystallization phenomenon is absent. In that case we find also, 
as expected, a dependence of the local structure on the glass history. 
This dependence is maximum when the intermediate temperature is close 
to $T_g$.\\
These results show that the Vorono\"\i\ cell characteristics 
permit to clearly distinguish recrystallization from ``true'' glassy 
state relaxations in the case of the soft-sphere glass. In the case of 
silica an extension of this work would be to check if the change of 
the local structure observed close to $T_g$ is detectable in other quantities.\\
{\bf Acknowledgements:}\\
It is not often that one can thank the referees for enlighting comments and
suggestions about the physics of a paper. This was our experience for this
publication and we want to stress it here. Therefore it is a pleasure
to thank also Pr. A. Fuchs for his editorial work on this article.


\newpage

\vspace*{-20mm}
\begin{figure}
\psfig{file=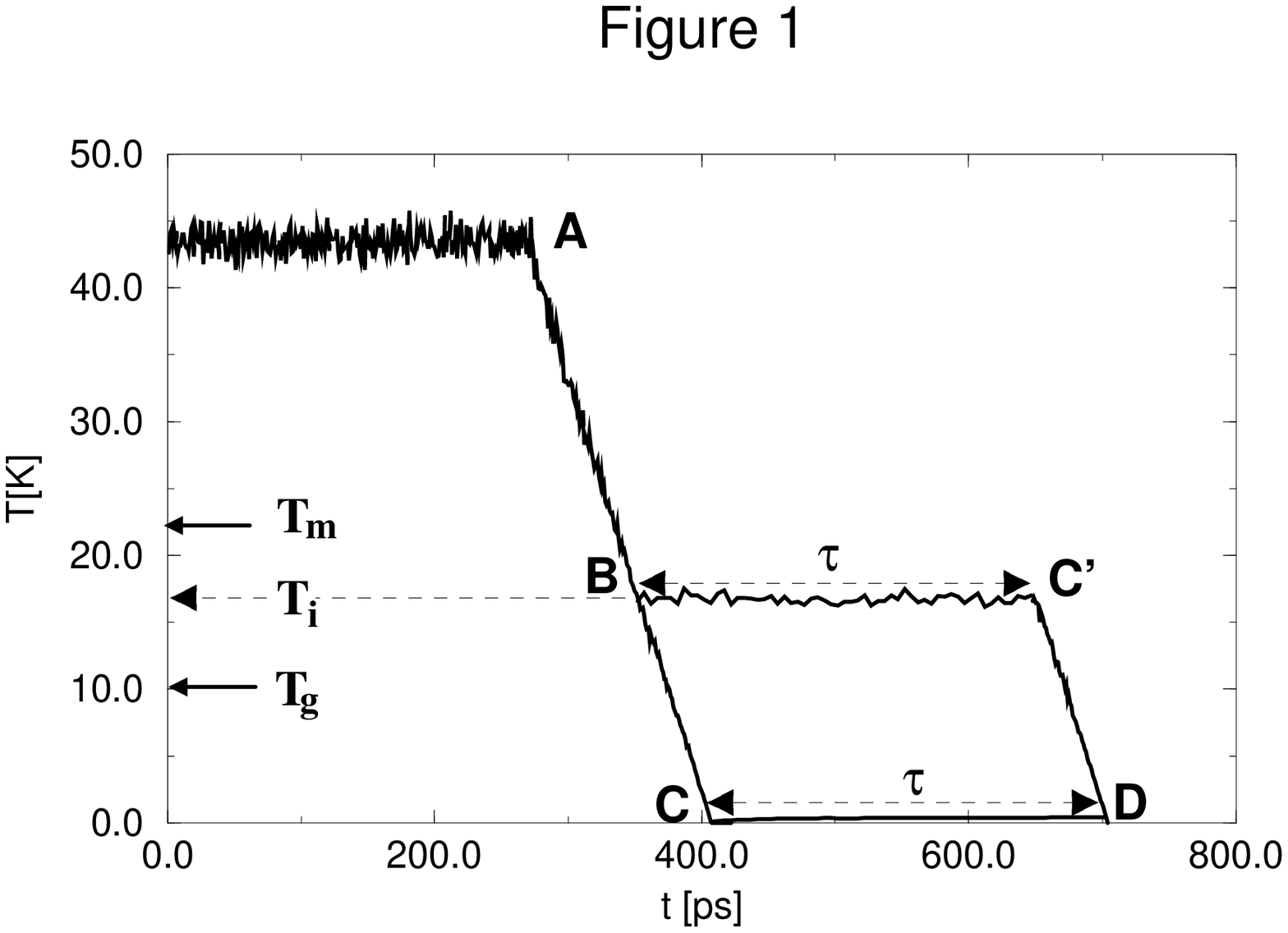,width=8.5cm,height=6.5cm}
\caption{Example of the quenching procedure used in this 
study: a relaxed liquid (point A) is quenched to 0K following either 
the path ACD or the path ABC'D.}
\label{Fig. 1}
\end{figure}

\vspace*{-15mm}
\par\noindent
\begin{figure}
\psfig{file=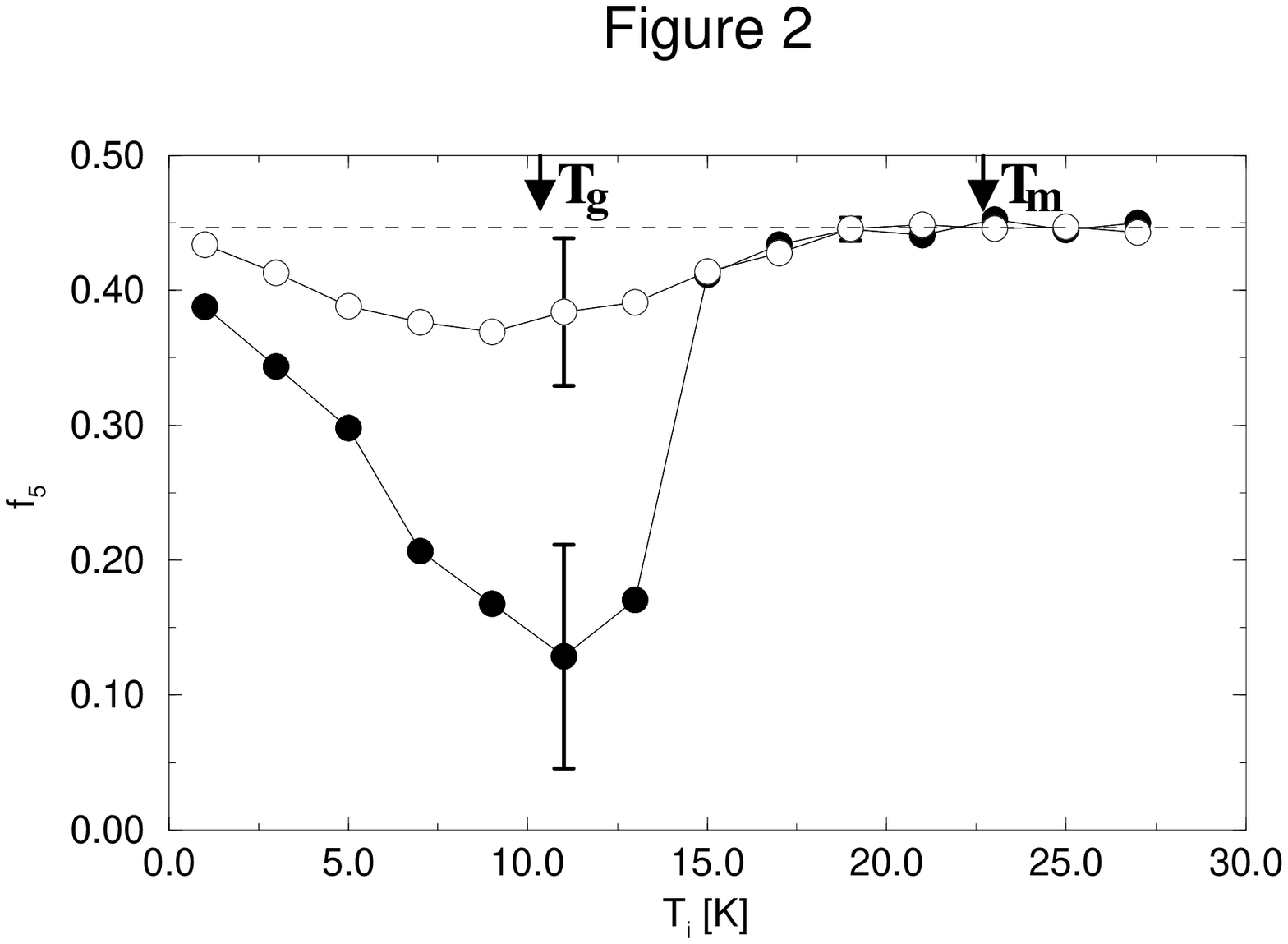,width=8.5cm,height=6.5cm}
\caption{Variation of the fraction of pentagonal faces,
$f_5$, as a function of the intermediate temperature $T_i$ for
$\tau = 100$ ps ( $\circ$) and $\tau = 300$ps ($\bullet$). The horizontal
dashed line is a guide for the eye and indicates the value of $f_5$ in
the reference structure at 0K.}
\label{Fig. 2}
\end{figure}

\vspace*{-15mm}
\par\noindent
\begin{figure}
\psfig{file=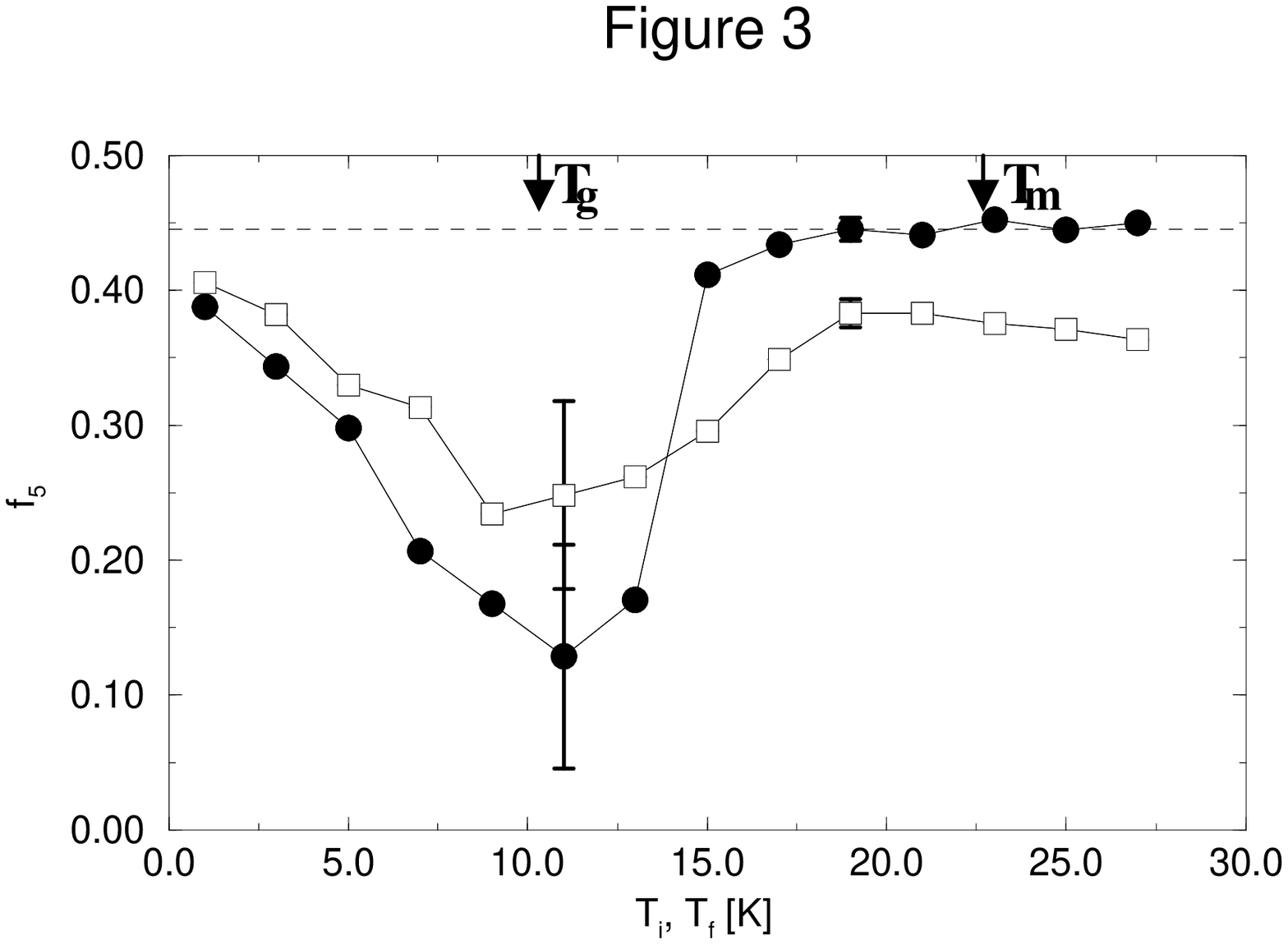,width=8.5cm,height=6.5cm}
\caption{$f_5$ as a function of $T_i$ ($\bullet$) and
$T_f$ ($\Box$). $T_f$ is the final temperature after a relaxation time 
$\tau = 300$ps (temperature at point C' in Fig. 1). The horizontal
dashed line is a guide for the eye and indicates the value of $f_5$ in
the reference structure at 0K.}
\label{Fig. 3}
\end{figure}

\vspace*{-20mm}
\par\noindent
\begin{figure}
\psfig{file=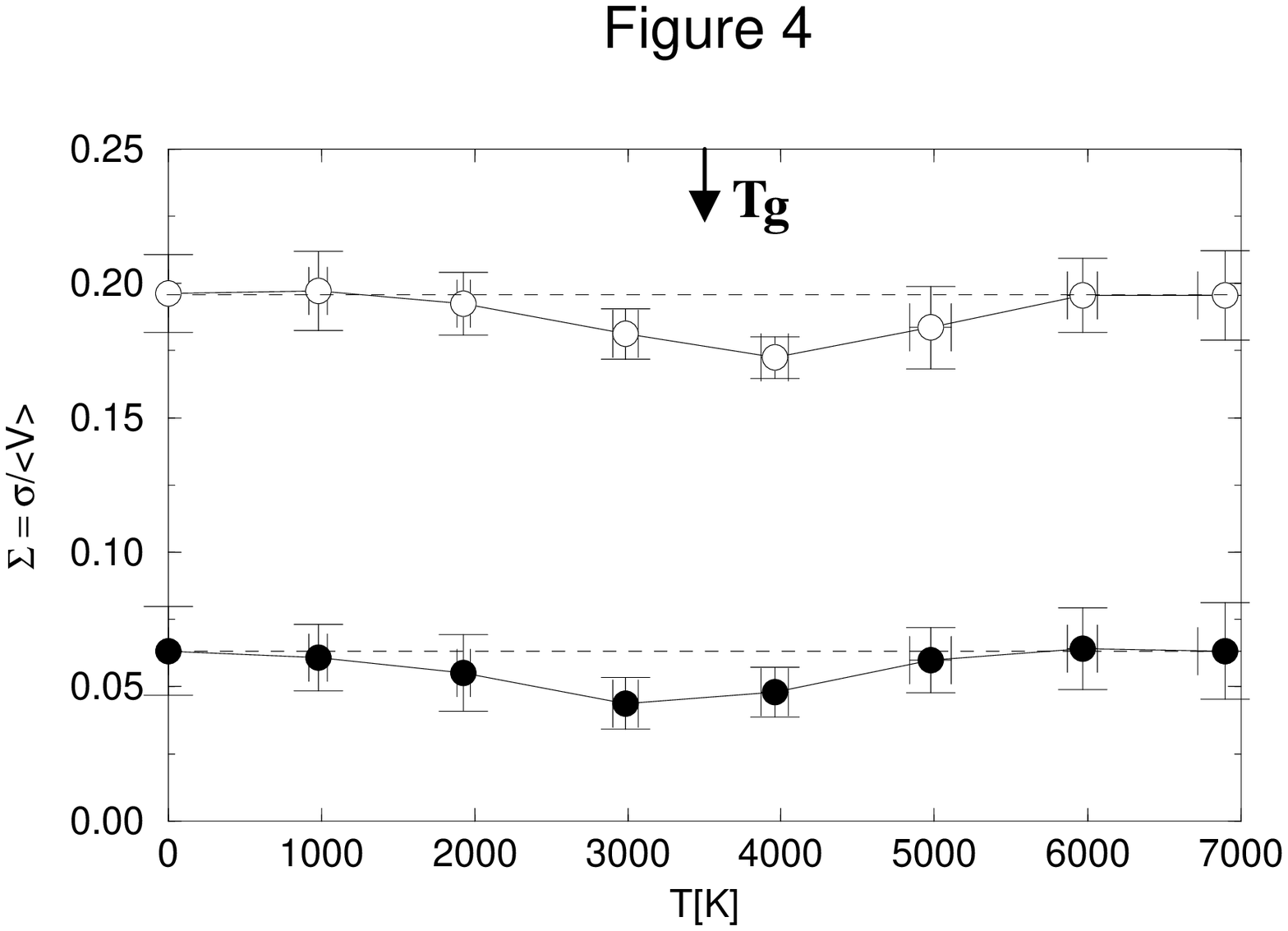,width=8.5cm,height=6.5cm}
\caption{Variation of the reduced standard deviations 
$\Sigma = \sigma/<V>$, for the oxygen ($\circ$) and the silicon ($\bullet$) 
atoms in our silica system, as a function of the intermediate 
temperature $T_i$ for $\tau$ = 42 ps. The horizontal lines show 
the corresponding values in the reference structure at 0K and are 
guides for the eye.}
\label{Fig. 4}
\end{figure}


\end{document}